\documentclass[aps, prd, twocolumn, lengthcheck, superscriptaddress, 
nofootinbib]{revtex4-1}

\usepackage{xcolor}
\usepackage{color}
\usepackage{hyperref}
\usepackage{ulem}
\usepackage{graphicx}
\usepackage{mathrsfs}
\usepackage{amsmath}
\usepackage{slashed}
\usepackage{threeparttable}
\usepackage{subcaption}

\def\be{\begin{eqnarray}}
\def\ee{\end{eqnarray}}

\newcommand*{\nm}{\nonumber}

\begin{document}
\title{Composition of scalar mesons and their effects on nuclear matter properties in an extended linear sigma model}

\author{Yao Ma}
\email{mayao@nju.edu.cn}
\affiliation{School of Frontier Sciences, Nanjing University, Suzhou, 215163, China}

\author{Yong-Liang Ma}
\email{ylma@nju.edu.cn}
\affiliation{School of Frontier Sciences, Nanjing University, Suzhou, 215163, China}
\affiliation{International Center for Theoretical Physics Asia-Pacific (ICTP-AP) , UCAS, Beijing, 100190, China}

\begin{abstract}
It has been argued that the isoscalar and isovector mesons play significant roles in nuclear matter and neutron star structures. We improve the extended linear sigma model with baryons, proposed in our previous work, by introducing the flavor structures constructed from antisymmetric tensors of chiral representations to study these physics. The parameter space of this model is refined with well-reproduced nuclear matter properties at saturation density by the lowest-order Lagrangian, ensuring consistency with vacuum results, such as \(f_\pi \approx 134 \, \text{MeV}\).
The anticipated plateaulike behaviors of the symmetry energy are predicted at intermediate densities, which is crucial for the consistency of GW170817 and the neutron skin thickness of \(\text{Pb}^{208}\).
Subsequently, neutron star structures are calculated using several parameter sets, and the results for the nuclear matter properties at saturation density align with empirical values.
It is found that the neutron star structures are sensitive to the couplings between the isovector \(a_0\) meson and nucleons and the four-vector meson couplings: small values of both are favorable.
Meanwhile, nuclear matter properties at saturation density favor larger values of the latter and are not sensitive to the former.
This signifies the statistical significance of neutron star observations when obtaining realistic chiral effective field theories or models at various densities.
The parameter set favored by neutron star observations also aligns the behavior of the sound velocity with the conformal limit at high densities relevant to cores of massive stars.
It is hoped that the results of this work can guide future studies on the relationship between the microscopic symmetry of strong interactions and macroscopic phenomena.

\end{abstract}

\maketitle

\allowdisplaybreaks{}
\section{Introduction}

Nuclear matter (NM), a system dominated by the nonperturbative QCD, has been studied for several decades. But there are still great uncertainties about its properties, especially at compact star densities, due to our little knowledge on the phase structure of strong interaction matter~\cite{Fukushima:2010bq,Lattimer:2015nhk,Chen:2017och,Baym:2017whm,Gil:2018yah,Ma:2019ery,Li:2019xxz,Sorensen:2023zkk}.
The properties of NM are highly sensitive to details of nucleon interactions. Therefore, a reliable model of NM is not only significant for revealing neutron star (NS) structures~\cite{Lattimer:2000nx,Ozel:2016oaf}, but also can shed a light on characters of the hadrons involved.


In modeling NM, based on the phenomenological one-boson exchange for nucleon interactions~\cite{Holinde:1975vg, Erkelenz:1974uj, Nagels:1977ze, Machleidt:1989tm}, it has long been recognized that the scalar meson resonances, such as $\sigma, f_{0}$ and $a_0$, play indispensable roles by using the mean field approach (MFA)~\cite{Walecka:1974qa,Serot:1984ey,Serot:1997xg}.~\footnote{Hereafter, following the Particle Data Group, we denote the isovector scalar mesons as $a_0$.}
Without the isoscalar scalar mesons $\sigma/f_{0}$, which provide the attractive force between nucleons, it's hard to obtain the empirical values of the saturation density $n_0 \simeq 0.16$~fm$^{-3}$ and binding energy $E_0 \simeq 16.0$~MeV~\cite{Holinde:1975vg, Erkelenz:1974uj, Nagels:1977ze, Machleidt:1989tm}.
And the self-interactions of the $\sigma$ meson are necessary to reduce the nuclear incompressibility coefficient of symmetric nuclear matter~\cite{Boguta:1977xi} and other empirical values~\cite{Motohiro:2015taa}.
In addition to the isoscalar scalar mesons, the isovector scalar mesons are important for describing the asymmetric nuclear matter and hence the properties of neutron stars~\cite{Kubis:1997ew,Hofmann:2000vz,Liu:2001iz,Chen:2007ih,Roca-Maza:2011alv,Wang:2014jmr,Typel:2020ozc,Kong:2025dwl}.
Furthermore, it is found that the scalar-isoscalar scalar meson couplings, e.g., $\sigma a_0^2$ and $\sigma^2 a_0^2$ have large effects on both the symmetry energy $E_{\rm sym}$ and its slope $L$~\cite{Zabari:2018tjk,Kubis:2020ysv,Miyatsu:2022wuy,Li:2022okx}.
Explicitly, they soften $E_{\rm sym}$ at intermediate densities but stiffen $E_{\rm sym}$ at high densities.

Normally, in the models of NM, the scalar mesons are included with respect to that quantum numbers, such as parity, spin and isospin, less attention has been paid to other properties related to QCD.
From a QCD point of view, the structures of scalar mesons are still under debate, especially the light ones (see, e.g., Ref.~\cite{Close:2002zu} for a review).
From the mass ordering of the scalar mesons below $1$~GeV, it is difficult to regard them as the pure two-quark states but an admixture of the two- and four-quark configurations.
Considering that the two-quark and four-quark components of scalar mesons have different axial transformation properties, we recently proposed an extended linear sigma model (ELSM) with baryons, where baryons are taken as diquark-quark configurations, to study the properties of NM, and the model is noted as baryonic extended linear sigma model (bELSM).~\cite{Ma:2023eoz}.

In bELSM, the masses of the mesons and baryons are generated from the spontaneous chiral symmetry breaking.
Because of the conservation of axial transformation, only the two-quark component of the scalar mesons couples to the baryons at the lowest-order, which may provide deeper insights into the constituents of the scalar mesons from the NM and NS properties.
The $\rm U(1)_A$ anomaly is involved in the 't Hooft determinant~\cite{Rosenzweig:1979ay,Kawarabayashi:1980dp,Hsu:1998jd,Fariborz:2005gm}.
Another important feature of the bELSM for the phenomenological analysis is the inclusion of all possible types of mesons below \(1\ \rm GeV\) and hyperons which significantly affect NS structures~\cite{Gal:2016boi}.

After pinning down the NM properties at saturation density \(n_0\), a plateaulike structure of the symmetry energy, \(E_{\rm sym}\), at intermediate densities, \(\sim 2n_0\), was found in the bELSM.
This is consistent with the analysis based on various of approaches in Refs.~\cite{Paeng:2017qvp,Ma:2018xjw,Ma:2018jze,Ma:2021nuf,Lee:2021hrw,Zabari:2018tjk,Miyatsu:2022wuy,Li:2022okx}, and satisfies the constraints of \(\rm {}^{208}Pb\)~\cite{PREX:2021umo,Reed:2021nqk} and tidal deformations of NS from GW170817~\cite{LIGOScientific:2016aoc}.
We also investigated the effects of parameter variations on different coefficients of Taylor expansions of energy per nucleon, \(E=\mathcal{E}/n \), by density, \(n=n_n+n_p\), and asymmetry, \({\delta} = (n_n-n_p)/n\), to provide further guidance on the interaction parameterization of NM,
\begin{widetext}
\be
E(n,{\delta}) & \simeq & E(n_0,0) + E_{\rm sym}(n_0){\delta}^2 + L(n_0){\delta}^2\chi+\frac{K(n_0)}{2!} \chi^2+\frac{J(n_0)}{3!} \chi^3 + \mathcal{O}\left({\delta}^4\right)+\mathcal{O}\left(\chi^4\right)+\mathcal{O}\left({\delta}^2\chi^2\right)\nonumber\\
& \simeq & E_0 + \left.\frac{{\delta}^2}{2}\frac{\partial^2 E(n_0, {\delta})}{\partial {\delta}^2}\right|_{{\delta}=0}+{\delta}^2\chi\left.\left(3 n \frac{\partial E_{\mathrm{sym}}(n)}{\partial n}\right)\right|_{n=n_0}+\frac{\chi^2}{2!}\left.\left(9 n^2 \frac{\partial^2 E_0(n)}{\partial n^2}\right)\right|_{n=n_0} \nonumber\\
& & {} + \frac{\chi^3}{3!}\left.\left(27 n^3 \frac{\partial^3 E_0(n)}{\partial n^3}\right)\right|_{n=n_0}+\mathcal{O}\left({\delta}^4\right)+\mathcal{O}\left(\chi^4\right)+\mathcal{O}\left({\delta}^2\chi^2\right)\ ,
\ee
\end{widetext}
where \(\chi \equiv\left(n-n_0\right) / 3 n_0\).
\(L(n_0)\), \(K(n_0)\), and \(J(n_0)\) refer to the symmetry energy slope, incompressibility, and skewness of NM at \(n_0\), respectively.

However, when calculating the \(E_{\rm sym}(n)\) using bELSM, the plateaulike structure is found to be lower than the results in Refs.~\cite{,Miyatsu:2022wuy,Li:2022okx}, leading to a loose structure of NS with tidal deformation \(\Lambda_{1.4} > 1000\).
This is due to the couplings related to \(\rho\) and \(\omega\) mesons being constrained by the \(\rm U(3)_V\) symmetry, which is not entirely realistic, especially in the nuclear medium~\cite{Paeng:2011hy}.
Therefore, improving the bELSM to resolve these problems and achieve a better description of the NM properties and NS structures is the purpose of this work.

When constructing the interactions, especially in the baryonic sector, only operators like \({\rm Tr}(\bar{B}\gamma_{\mu}V^{\mu}B)\) were considered in Ref.~\cite{Ma:2023eoz}, which make the couplings \(g_{\rho NN}\) and \(g_{\omega NN}\) the same.
However, another type of operator, \(\epsilon_{abc}\epsilon^{def}\bar{B}_{ad}\gamma_{\mu}B_{be}V^{\mu}_{cf}\)~\cite{Papazoglou:1997uw}, which can split the couplings of the singlet and octet vector mesons to baryons.
In this work, we extend the bELSM by including this kind of operators in both the mesonic and baryonic sectors at the lowest-order and improve the parameter space of the original bELSM.
The 1two-quark terms, which were necessary to maintain reasonable behavior at high densities in the work of Ref.~\cite{Ma:2023eoz}, can be removed, enhancing the rationality of the power counting in the bELSM.
The vacuum expectation values of the scalar meson fields are close to the results of Ref.~\cite{Fariborz:2007ai}, leading to a reasonable \(f_{\pi}\) (quark condensate) and low-energy pion scattering results.

With the improved bELSM at the lowest-order, the NM properties at \(n_0\) and the bare masses of the mesons and baryons can be well reproduced.
The bELSM is then applied to calculating NS structures. It is found that NS structures are sensitive to the NM properties at \(n_0\), especially the incompressibility, \(K(n_0)\), which also affects the sound velocity (SV) behavior at various densities.
The relationship between NS and SV is very subtle, as detailed in Refs.~\cite{Bedaque:2014sqa,zhang2024peaksoundvelocityscale}. When \(K(n_0)\) falls within empirical constraints, the SV softens significantly, causing the maximum mass of NS to be lower than the observed value, \(2.14\pm 0.10 M_{\odot}\)~\cite{NANOGrav:2019jur}.
To meet the constraints of the maximum mass of NS and the mass-radius (M-R) relation of NS from GW170817~\cite{LIGOScientific:2016aoc}, \(K(n_0)\) should be larger than the empirical constraints, approximately \(500\ \rm MeV\).
This \(K(n_0)\) also results in an interesting behavior of SV, approaching the conformal limit at high densities, \(n\sim 5 n_0\), whereas other SVs from smaller \(K(n_0)\) will be too soft.
This highlights the statistical significance of NS observations when obtaining realistic chiral effective field theories or models~\cite{Guo:2023mhf}.
In addition, the influence of portions in the two-quark and four-quark constituents of the scalar mesons on NS structures and NM properties is also shown and their physical magnitudes are estimated. These discussions set up a relationship between microscopic symmetry and macroscopic phenomena.

The rest of this article is organized as follows: In Sec.~\ref{sec:ELSM}, we first introduce the improved bELSM at the lowest-order under the relativistic mean field (RMF) approximation.
Then we study the properties of the scalar mesons on the NM properties and NS structures in detail.
In Sec.~\ref{sec:DO}, the discussion on the results of the current work and the outlook are provided. The details of the model construction and approximation are given in the Appendix.

{\section{Extended linear sigma model at finite densities}}
\label{sec:ELSM}

\subsection{Theoretical framework}

In this work, we consider the lowest-order bELSM given in Ref.~\cite{Ma:2023eoz} including terms up to eight quark lines with a single trace and operators of dimension 4. In addition, as mentioned in the Introduction, we consider the terms with three-rank Levi-Civita tensor in flavor space. The Lagrangian contributing under RMF approximation is given by
\be
\mathcal{L}=\mathcal{L}_{\rm B} + \mathcal{L}_{\rm M} + \mathcal{L}_{\rm V}
\label{eq:LagMF}
\ee
where
\begin{subequations}
\be
\mathcal{L}_{\rm B} & = & {\rm Tr}\left(\bar{B}i\slashed{\partial}B\right)+c{\rm Tr}\left(\bar{B}V\gamma^0 B\right) \nonumber\\
& &{} -g{\rm Tr}\left(\bar{B}S' B\right)+h\epsilon_{abc}\epsilon^{def}\bar{B}_{ad}\gamma^{0}B_{be}V_{cf}\nonumber\\
& &{}-e\epsilon_{abc}\epsilon_{def}\bar{B}_{ad}B_{be}S_{cf}'\ , \label{eq:LB}\\ 
\mathcal{L}_{\rm M} & = & c_2{\rm Tr}S^{\prime 2}-d_2{\rm Tr}\hat{S}^{\prime 2}-c_4{\rm Tr}S^{\prime 4} \nonumber\\
& &{} -2e_3\epsilon_{abc}\epsilon_{def}S_{ad}'S_{be}'\hat{S}_{cf}'\ ,\label{eq:Vm}\\
\mathcal{L}_{\rm V} & = & \tilde{h}_2{\rm Tr}\left(S^{\prime2}V^2\right)+\tilde{g}_3{\rm Tr}V^4 \nonumber\\
& &{} +a_1\epsilon_{abc}\epsilon_{def}V_{ad}V_{be}\left(S^{\prime 2}\right)_{cf}\ .\label{eq:Vv}
\ee
\end{subequations}
The $h,\ e$ and $a_1$ terms are the terms which were not considered in Ref.~\cite{Ma:2023eoz}.
The details of the model construction are given in Appendix~\ref{app:LbELSM}.

In Lagrangian~\eqref{eq:LagMF}, $S^\prime$ is the two-quark configuration of scalar states which can be decomposed as
\be
S^\prime& = & \frac{1}{\sqrt{2}}(\lambda_8 f_0^\prime+\lambda_3 a_0^\prime)+\frac{1}{\sqrt{3}}I\sigma^\prime,
\ee
with \(\lambda_3\) and \(\lambda_8\) being the third and eighth components of Gell-Mann matrices and \(I\) being the identity matrix. The similar decomposition applies to the four-quark configuration \(\hat{S}^\prime\). $V$ is the zero component of vector fields which survives in the MFA 
\be
V & = & \frac{1}{2}\left(\lambda_3\rho+\frac{1}{\sqrt{3}}\lambda_8\omega\right) + \frac{1}{3}I\omega. 
\ee
When restricting to the two-flavor case, the baryon octet is reduced to
\begin{equation}
	B \rightarrow\left(\begin{array}{ccc}
		0 & 0 & p \\
		0 & 0 & n \\
		0 & 0 & 0
	\end{array}\right)\ .
\end{equation}

So, with \(p\) and \(n\), there are isoscalar scalar configurations $\sigma^\prime/\hat{\sigma}^\prime$, $f_0^\prime/\hat{f}_0^\prime$ and isovector scalar configurations $a_0^\prime/\hat{a}_0^\prime$ in model~\eqref{eq:LagMF}. After chiral symmetry breaking, the mixing of these scalar configurations yields the physical states of scalar mesons. The spontaneous chiral symmetry breaking is achieved by the vacuum expectation values of both two-quark and four-quark fields: \(\langle\sigma'\rangle=\sqrt{3}\alpha\) and \(\langle\hat{\sigma}'\rangle=\sqrt{3}\beta\), which are determined by the minimum of the potential \(\mathcal{L}_{\rm M}\),
\be
\left\langle\frac{\partial \mathcal{L}_{\rm M}}{\partial \sigma'}\right\rangle & = & 2 \alpha\left(-c_2+2 c_4 \alpha^2+4 e_3 \beta\right)=0\ , \nonumber\\
\left\langle\frac{\partial \mathcal{L}_{\rm M}}{\partial \hat{\sigma}'}\right\rangle & = & 2\left(d_2 \beta+2 e_3 \alpha^2\right)=0\ ,
\label{eq:VEV}
\ee
where the three-flavor symmetry is used. The physical states \(\sigma\), \(a_0\), and \(f_0\) are obtained by the mixing of the two-quark and four-quark configurations, via
\be
& & \sigma=\cos\theta_0\sigma'+\sin\theta_0\hat{\sigma}'\ , \nonumber\\
& & a_0=\cos\theta_8 a_0'+\sin\theta_8\hat{a_0}'\ , \nonumber\\
& & f_0=\cos\theta_8 f_0'+\sin\theta_8\hat{f_0}'\ ,
\label{eq:mixing}
\ee
which are determined by the diagonalization of the mass matrix of the scalar mesons.
As one can see, there are two sets of physical scalar mesons: one contains the lowest-lying scalar mesons \(\sigma\), \(a_0\), and \(f_0\), while the other set refers to the excitation states in the corresponding channels, whose contribution to the equation of state (EOS) of NM is negligible but their spectra are considered to settle down the free parameters. 

\subsection{Nuclear matter properties and neutron star structures}
\label{sec:NM}

Using the relations in Eq.~\eqref{eq:VEV}, the degrees of freedom of the parameter space are reduced to \(11\), explicitly, \(\alpha\), \(\beta\), \(c_4\), \(e_3\), \(\tilde{h}_2\), \(\tilde{g}_3\), \(a_1\), \(c\), \(g\), \(h\), and \(e\).
The physical quantities that can be used to constrain these parameters are the bare masses of the scalar mesons and nucleons, as well as the NM properties around saturation density, \(n_0\). The empirical constraints and corresponding fitted values are shown in Tables~\ref{tab:mass} and~\ref{tab:nm}, and the fitted parameters are listed in Table~\ref{tab:para}. 

\begin{widetext}
\begin{table*}[htbp]
	\caption{
        Values of the parameters used in this work.
        $\alpha$ and $\beta$ are in units of MeV, and $e_3$ is in units of GeV.
        The other parameters are dimensionless.
        The parameter sets are denoted as el-g30g, el-g30e, el-g30eg, el-g350eg, el-g3100eg, and el-g3150eg, respectively; "el" refers to the bELSM; "g3" refers to the four-vector meson coupling in Eq.~\eqref{eq:Vv}, with the number following it denoting the corresponding magnitude; "e" or "g" indicates that the coupling between baryons and scalar mesons is mainly from "e" or "g" terms in Eq.~\eqref{eq:LB}, and "eg" indicates that the contributions of the "e" and "g" terms are similar.
	}
	\label{tab:para}
	\begin{threeparttable}   
		\begin{tabular}{@{}ccccccccccccc}
			\hline
			\hline
			& $\alpha$ & $\beta$& $e_3$ & $c_4$ & $\tilde{h}_2$  & $\tilde{g}_3$ & $c$ & \(g\) & \(a_1\) & \(h\) & \(e\)\\
			\hline
			el-g30g & $61.1$  & $24.7$ & $-2.06$  & $44.0$ &  $80.1$ & $1.59$ & $-0.792$ & $15.4$& $4.25$ & $11.4$ & $-0.0270$ \\
            \hline
			el-g30e & $61.1$  & $24.4$ & $-2.05$  & $43.6$ &  $80.0$ & $0.542$ & $11.4$ & $0.234$& $4.17$ & $-0.790$ & $15.1$ \\
            \hline
			el-g30eg & $61.4$  & $26.4$ & $-2.10$  & $45.6$ &  $79.3$ & $0.397$ & $9.51$ & $6.54$& $4.10$ & $-2.61$ & $8.75$ \\
            \hline
			el-g350eg & $61.2$  & $25.6$ & $-2.10$  & $44.4$ &  $79.9$ & $51.5$ & $10.1$ & $6.35$& $4.14$ & $-2.65$ & $9.00$ \\
            \hline
			el-g3100eg & $60.8$  & $24.0$ & $-2.09$  & $42.4$ &  $80.8$ & $100$ & $10.6$ & $7.10$& $4.19$ & $-2.88$ & $8.34$ \\
            \hline
			el-g3150eg & $60.7$  & $24.3$ & $-2.11$  & $42.0$ &  $81.0$ & $150$ & $11.1$ & $7.15$& $4.18$ & $-3.05$ & $8.30$ \\
			\hline
			\hline
		\end{tabular}
	\end{threeparttable}
\end{table*}

\begin{table*}[htb]
	\caption{
        The spectra of hadrons are in units of $\mathrm{MeV}$.
        The empirical values are chosen as the real parts of the corresponding resonance T-matrix poles or masses from Ref.~\cite{ParticleDataGroup:2024cfk}.
        The excited state $f_0'$ is chosen to be $f_0(1370)$ with $a_0'$ being $a_0(1450)$ and $\sigma'$ being $f_0(1500)$.
        The constraints of $f_0(1370)$ and $a_0(1450)$ are combined together, due to \(\rm U(3)_V\) symmetry.
        The definition of the parameter set is shown in Table~\ref{tab:para}.
	}
    \label{tab:mass}
	\begin{threeparttable}
		\begin{tabular}{@{}cccccccc}
			\hline
			\hline
			& $m_N$ & $m_{\sigma}$ & $m_{f_0(a_0)}$ & $m_{f_0'(a_0')}$ & $m_{\sigma'}$ & $m_{\rho}$ & $m_{\rho}$\\
			\hline
			Empirical & $938$-$940$ & $400$-$800$ & $960$-$1010$ & $1250$-$1500$ & $1430$-$1530$ & $761$-$765$ & $782$-$783$\\
			\hline
			el-g30g & $939$ & $522$ & $989$ & $1480$ & $1510$ & $763$ & $783$\\
            \hline
			el-g30e & $939$ & $525$ & $991$ & $1480$ & $1510$ & $763$ & $783$\\
            \hline
			el-g30eg & $939$ & $498$ & $983$ & $1500$ & $1520$ & $763$ & $783$\\
            \hline
			el-g350eg & $939$ & $485$ & $991$ & $1470$ & $1510$ & $763$ & $783$\\
            \hline
			el-g3100eg & $939$ & $502$ & $994$ & $1470$ & $1510$ & $763$ & $783$\\
            \hline
			el-g3150eg & $939$ & $485$ & $991$ & $1470$ & $1510$ & $763$ & $783$\\
			\hline
			\hline
		\end{tabular}
	\end{threeparttable}
\end{table*}

\begin{table*}[htb]
	\caption{
        Nuclear matter properties at saturation density \(n_0\).
        The empirical values and their error bars are set according to analyses from Refs.~\cite{Sedrakian:2022ata,Piekarewicz:2003br,Colo:2004mj,Khan:2012ps,Dutra:2012mb,Dutra:2014qga,Oertel:2016bki,Tews:2016jhi,Wang:2018hsw,Xie:2019sqb,Choi:2020eun,Grams:2022bbq}.
        The \(n_0\) is in units of \(\rm fm^{-3}\) and the other properties are in units of \(\rm MeV\).
        The definition of the parameter set is shown in Table~\ref{tab:para}.
    }
    \label{tab:nm}
	\begin{threeparttable}
		\begin{tabular}{@{}ccccccc}
			\hline
			\hline
			& $n_0$ & \(E_0\) & $E_{\mathrm{sym}}(n_0)$ & $J(n_0)$ & $L(n_0)$ & $K(n_0)$ \\
			\hline
			Empirical & $0.155\pm0.050$ & $-16.0\pm1.0$ & $31.7\pm3.2$ & $-200\pm600$ & $58.7\pm28.1$ & $250\pm50$ \\
			\hline
			el-g30g & $0.155$ & $-14.6$ & $30.9$ & $451$ & $83.6$ & $418$ \\
            \hline
			el-g30e & $0.155$ & $-14.6$ & $31.6$ & $479$ & $85.8$ & $419$ \\
            \hline
			el-g30eg & $0.155$ & $-14.6$ & $30.1$ & $421$ & $82.2$ & $415$ \\
            \hline
			el-g350eg & $0.155$ & $-15.2$ & $30.9$ & $-392$ & $80.7$ & $370$ \\
            \hline
			el-g3100eg & $0.155$ & $-15.4$ & $31.4$ & $-1020$ & $71.7$ & $317$ \\
            \hline
			el-g3150eg & $0.155$ & $-15.6$ & $31.6$ & $-1470$ & $63.7$ & $253$ \\
			\hline
			\hline
		\end{tabular}
	\end{threeparttable}
\end{table*}
\end{widetext}

It can be seen from Table~\ref{tab:mass} that all the parameter sets reproduce hadron spectra consistent with the results from the Particle Data Group~\cite{ParticleDataGroup:2024cfk}.
Regarding the NM properties at \(n_0\) shown in Table~\ref{tab:nm}, the saturation density is consistently fixed at \(0.155~\rm fm^{-3}\), and \(E_{\rm sym}\) aligns with the empirical value of \((30.9\pm1.9)~\rm MeV\). However, the other properties decrease with the magnitude of \(\tilde{g}_3\).
It appears that \(\tilde{g}_3\) should be approximately \(150\) to satisfy the constraints of NM properties.

\begin{figure}
    \centering
    \begin{subfigure}[]{0.35\textwidth}
        \includegraphics[width=\textwidth]{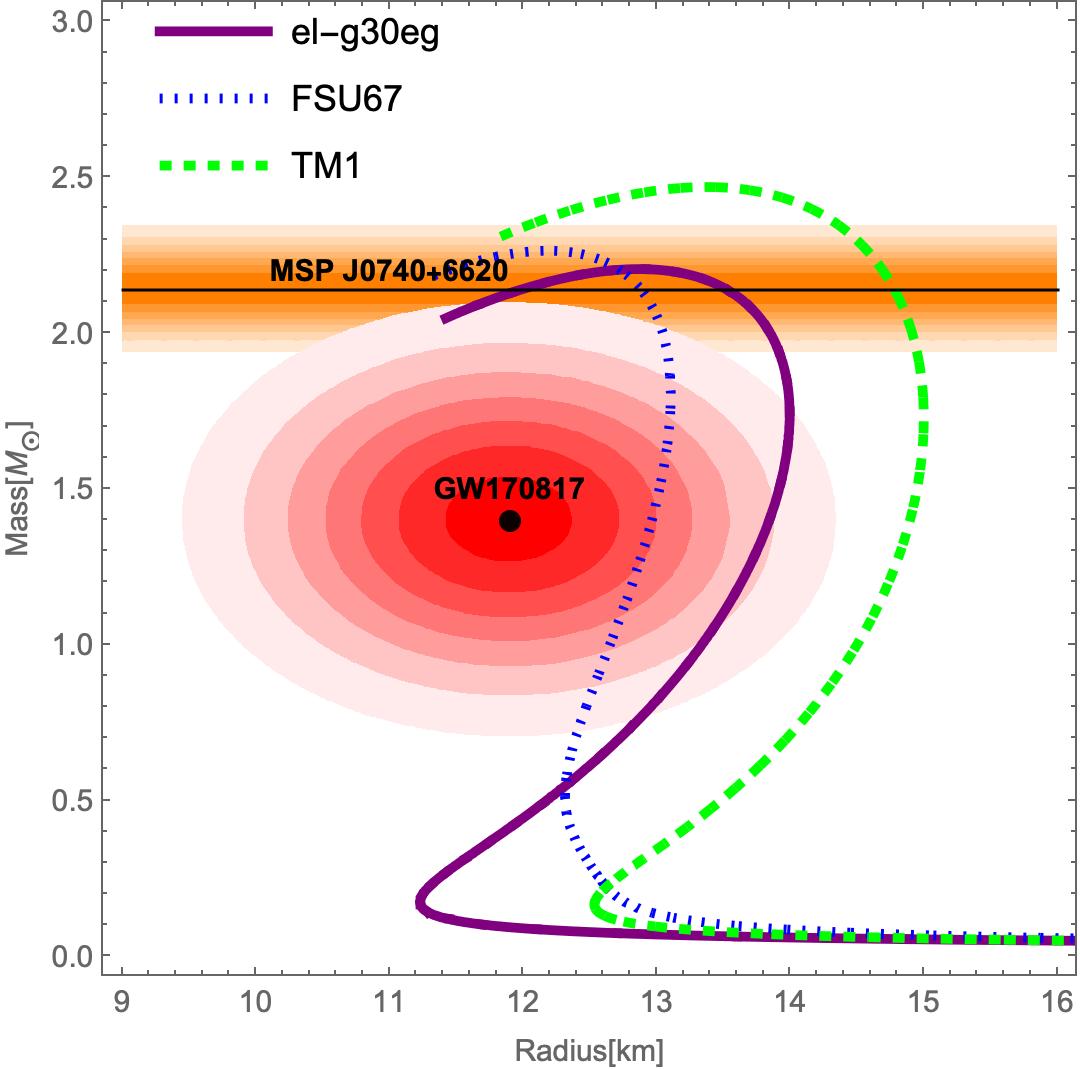}
        \caption{
            Comparison among TM1~\cite{Sugahara:1993wz}, FSU-\(\delta 6.7\)~\cite{Li:2022okx}, and el-g30eg.
        }
        \label{fig:MRsub1}
    \end{subfigure}
    \begin{subfigure}[]{0.35\textwidth}
        \includegraphics[width=\textwidth]{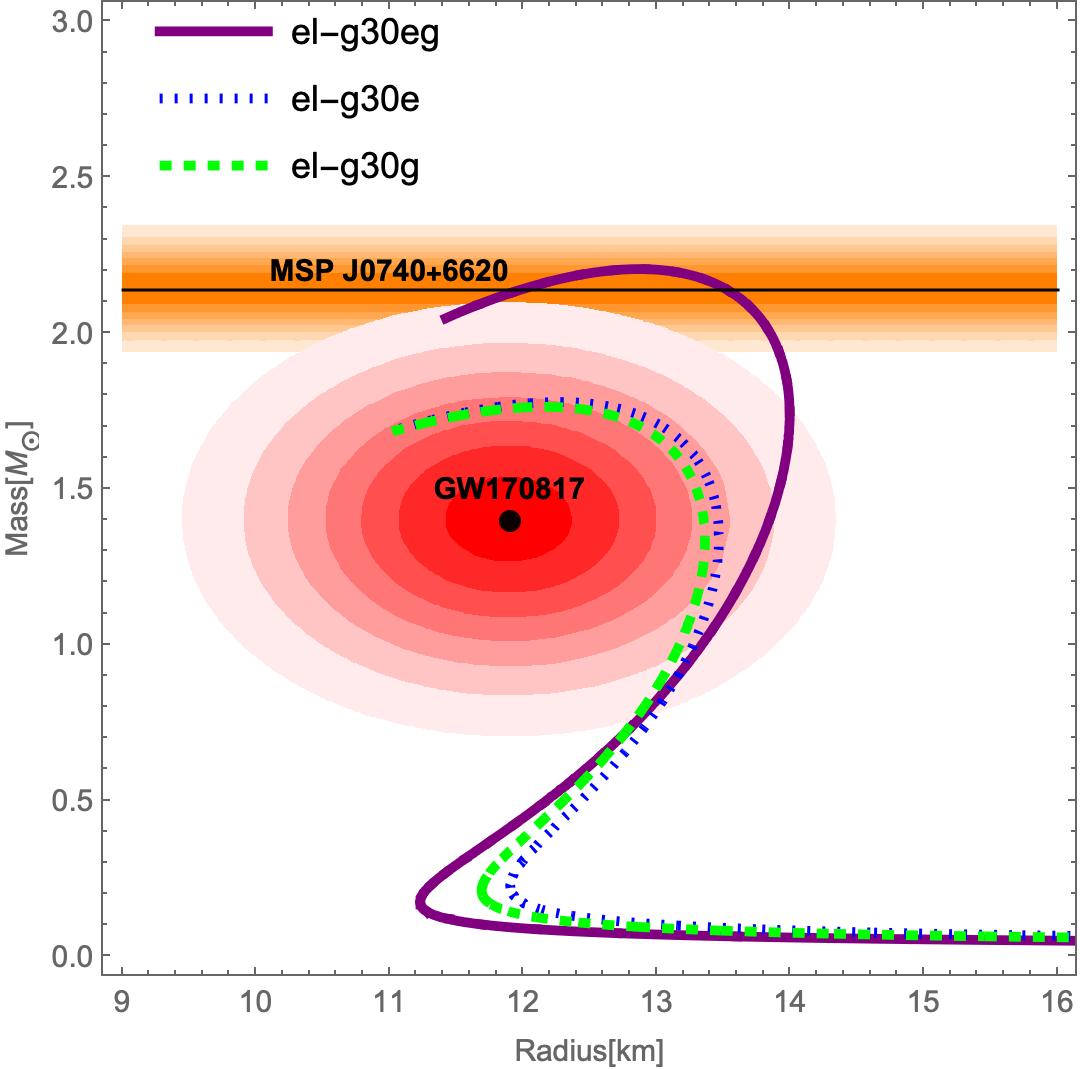}
        \caption{Comparison among cases with different magnitudes of \(e\) and \(g\) flavor structures.}
        \label{fig:MRsub2}
    \end{subfigure}
    \begin{subfigure}[]{0.35\textwidth}
        \includegraphics[width=\textwidth]{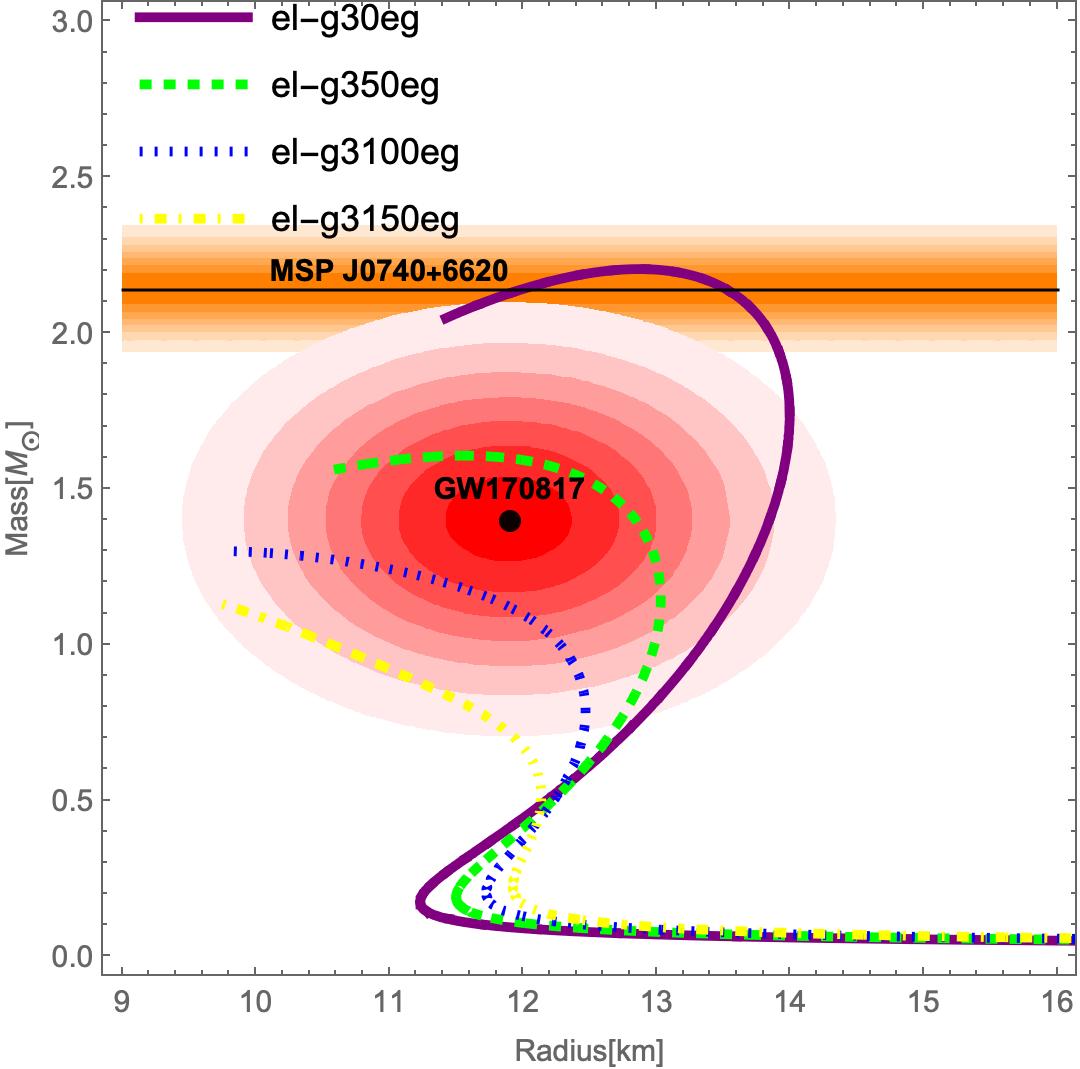}
        \caption{Compasrion among cases with different magnitude of \(\tilde{g}_3\) terms.}
        \label{fig:MRsub3}
    \end{subfigure}
    \caption{
        The M-R relation of NSs.
        The constraints MSP J0740+6620 are from Ref.~\cite{NANOGrav:2019jur} and GW170817 are from Ref.~\cite{LIGOScientific:2018cki}. Both are at \(95\%\) confidence level.
    }
    \label{fig:MR}
\end{figure}

When we extend our study beyond the saturation density to a broader density range to examine NS structures, these parameter sets result in significant differences.
The M-R relations of NSs are shown in Fig.~\ref{fig:MR}, which is computed using the Tolman-Oppenheimer-Volkoff equation~\cite{Tolman:1939jz,Oppenheimer:1939ne} with pure hadron EOS from different models.
As we can see in Fig.~\ref{fig:MRsub1}, the model without the \(a_0\) meson cannot reach the GW170817 constraint due to its overly loose structure of NSs.
The models with the \(a_0\) meson, both el-g30eg and FSU-\(\delta 6.7\), can satisfy the GW170817 and MSP J0740+6620 constraints simultaneously, highlighting the importance of the \(a_0\) meson---an isovector scalar meson in NS studies.
From Fig.~\ref{fig:MRsub2}, it is evident that only EOS with similar contributions from the \(e\) and \(g\) terms can meet the observational constraint of MSP J0740+6620; otherwise, the \(M_{\rm max}\) will be slightly lower.
However, the properties of NM at \(n_0\) are not sensitive to the contributions of the \(e\) and \(g\) terms. Regarding the \(\tilde{g}_3\) terms, as shown in Fig.~\ref{fig:MRsub3}, the \(M_{\rm max}\) of NSs is sensitive to the magnitude of \(\tilde{g}_3\): a larger \(\tilde{g}_3\) coupling---four vector meson couplings---leads to a smaller \(M_{\rm max}\) of NSs, even though the NM properties at \(n_0\) are more consistent with empirical values.

To further understand the effects of the meson-nucleon coupling on the NS structures, we combine the parameters in various of model to the meson-baryon couplings defined by
\be
\mathcal{L}_{\rm RMF}^{\rm OBE} & = & g_{\sigma NN}\bar{N}\sigma N + g_{a NN}\bar{N}a_0\tau_3 N\nonumber\\
& &{} + g_{f NN}\bar{N}f_0 N +g_{\omega NN}\bar{N}\omega\gamma_0 N\nonumber\\
& &{} +g_{\rho NN}\bar{N}\rho\tau_3\gamma_0 N,
\ee 
where \(N=\binom{p}{n}\) and \(\tau_3\) is the third component of Pauli matrix. The values of the couplings are list in Table~\ref{tab:g}.
\begin{table}[htb]\small
	\caption{
        The values of meson-baryon coupling in various of models.
    }
    \label{tab:g}
	\begin{threeparttable}
		\begin{tabular}{@{}ccccccc}
			\hline
			\hline
			& $g_{\sigma NN}$ & \(g_{{a_0 NN}}\) & $g_{fNN}$ & $g_{\omega NN}$ & $g_{\rho NN}$ \\
			\hline
			TM1~\cite{Sugahara:1993wz} & $-10.0$ & ---& --- & $-12.6$ & $-4.63$ \\
            \hline
			FSU-\(\delta\)6.7~\cite{Li:2022okx} & $10.2$ & $6.7$ & --- & $-13.4$ & $-7.27$\\
			\hline
			el-g30g & $-6.17$ & $5.13$ & $2.95$ & $-6.09$ & $5.30$\\
            \hline
			el-g30e & $-6.20$ & $-5.03$ & $3.00$ & $6.09$ & $5.30$\\
            \hline
			el-g30eg & $-5.97$ & $-0.671$ & $2.68$ & $6.06$ & $3.45$\\
            \hline
			el-g350eg & $-6.12$ & $-0.852$ & $2.85$ & $6.37$ & $3.71$ \\
            \hline
			el-g3100eg & $-6.36$ & $-0.442$ & $3.19$ & $6.73$ & $3.85$ \\
            \hline
			el-g3150eg & $-6.38$ & $-0.413$ & $3.20$ & $7.09$ & $4.04$ \\
			\hline
			\hline
		\end{tabular}
	\end{threeparttable}
\end{table}

In our bELSM, the magnitudes of the \(e\) and \(g\) terms determine \(g_{a NN}\), while \(g_{\sigma NN}\) and \(g_{f NN}\) remain almost the same across different parameter sets.
However, the M-R relation of NSs is sensitive to \(g_{a NN}\): a larger \(g_{a NN}\) leads to a smaller \(M_{\rm max}\) for NSs, as shown in Fig.~\ref{fig:MRsub2}.
This is not the case in FSU-\(\delta 6.7\), which does not consider $f_0$.
In bELSM, all possible terms involving the \({a_0}\) meson at the lowest-order are considered, including all three- and four-point interactions: self-couplings, couplings to vector mesons, and couplings to scalar mesons.
These terms become increasingly important with density, especially those relate to vector mesons.
By including all possible multimeson interactions at the lowest-order, respecting the chiral symmetry pattern at low energies, the \(g_{\sigma NN}\) and \(g_{\omega NN}\) at densities are slightly smaller than those in Walecka-type models~\cite{Holinde:1975vg, Erkelenz:1974uj, Nagels:1977ze, Machleidt:1989tm, Sugahara:1993wz, Li:2022okx}, with \(g_{\rho NN}\) remaining almost the same.

\begin{figure}
    \centering
    \begin{subfigure}[]{0.45\textwidth}
        \includegraphics[width=\textwidth]{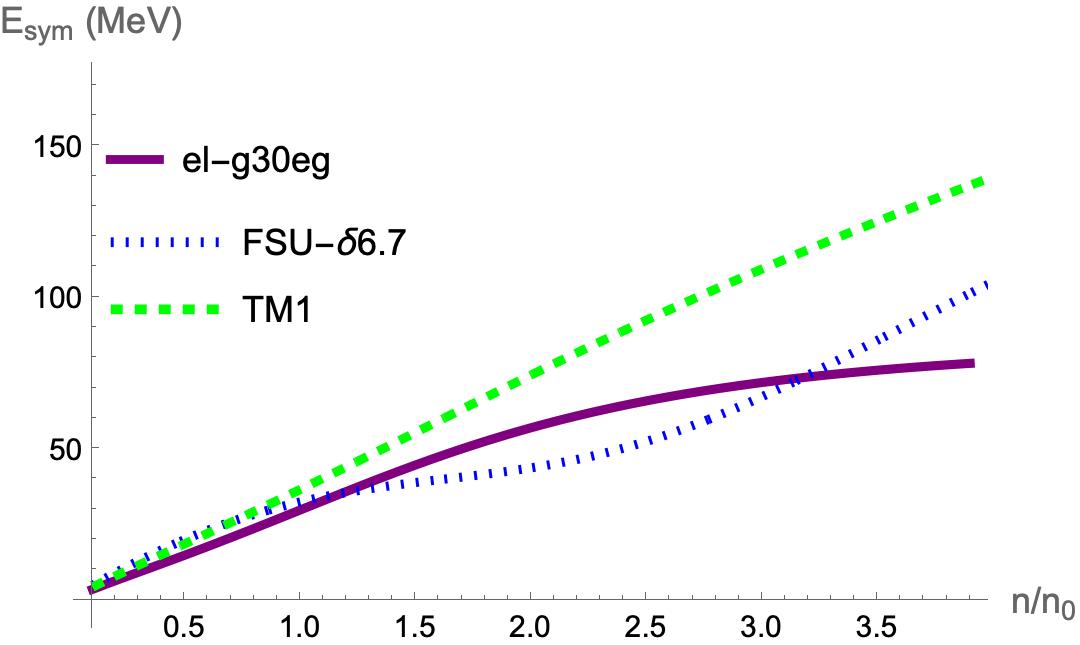}
        \caption{
            Comparison among TM1, FSU-\(\delta 6.7\), and el-g30eg.
            TM1 is from Ref.~\cite{Sugahara:1993wz}.
            FSU-\(\delta 6.7\) is from Ref.~\cite{Li:2022okx}.
        }
        \label{fig:Esymsub1}
    \end{subfigure}
    \begin{subfigure}[]{0.45\textwidth}
        \includegraphics[width=\textwidth]{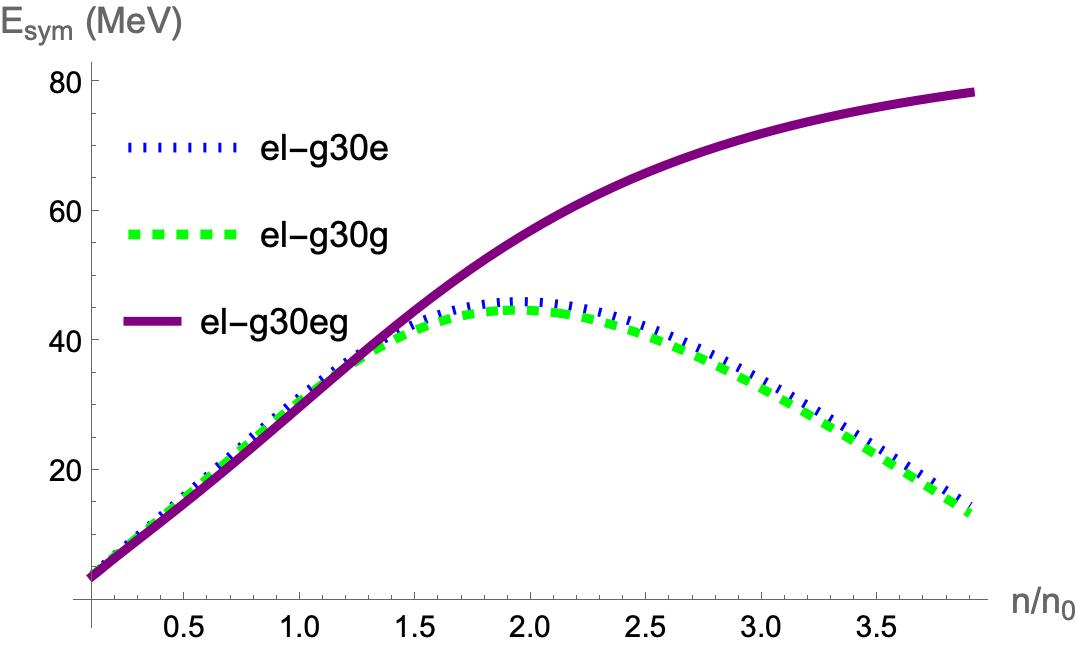}
        \caption{Comparison among cases with different magnitudes of \(e\) and \(g\) flavor structures.}
        \label{fig:Esymsub2}
    \end{subfigure}
    \begin{subfigure}[]{0.45\textwidth}
        \includegraphics[width=\textwidth]{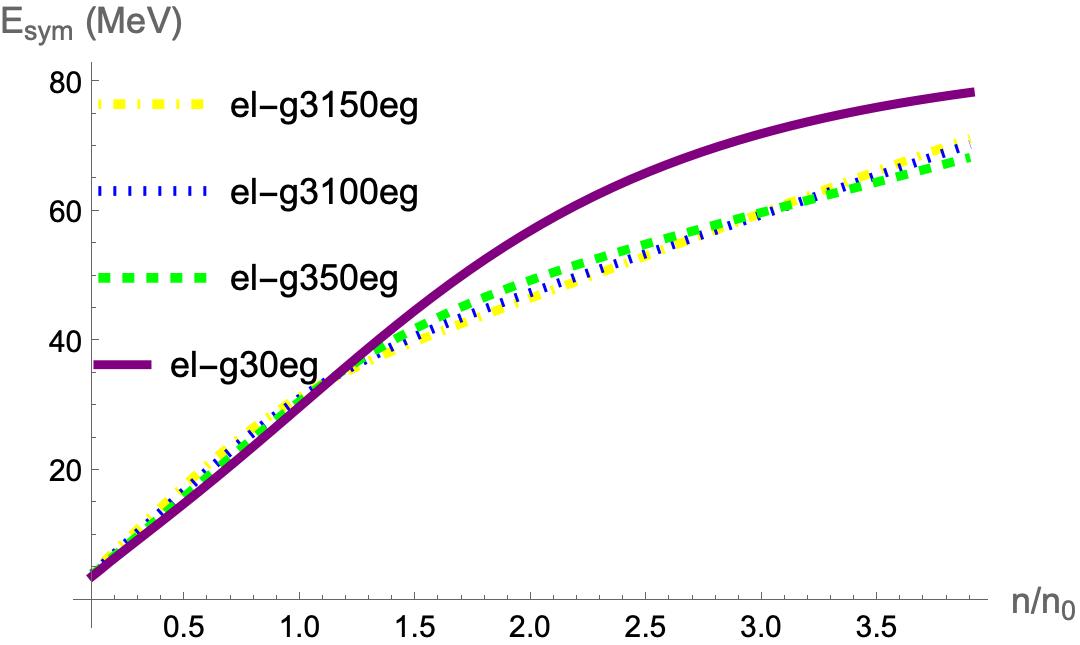}
         \caption{Comparison among cases with different magnitudes of \(\tilde{g}_3\) terms.}
        \label{fig:Esymsub3}
    \end{subfigure}
    \caption{
        The \(E_{\rm sym}(n)\) for different cases.
    }
    \label{fig:Esym}
\end{figure}

\begin{figure}
    \centering
    \begin{subfigure}[]{0.45\textwidth}
        \includegraphics[width=\textwidth]{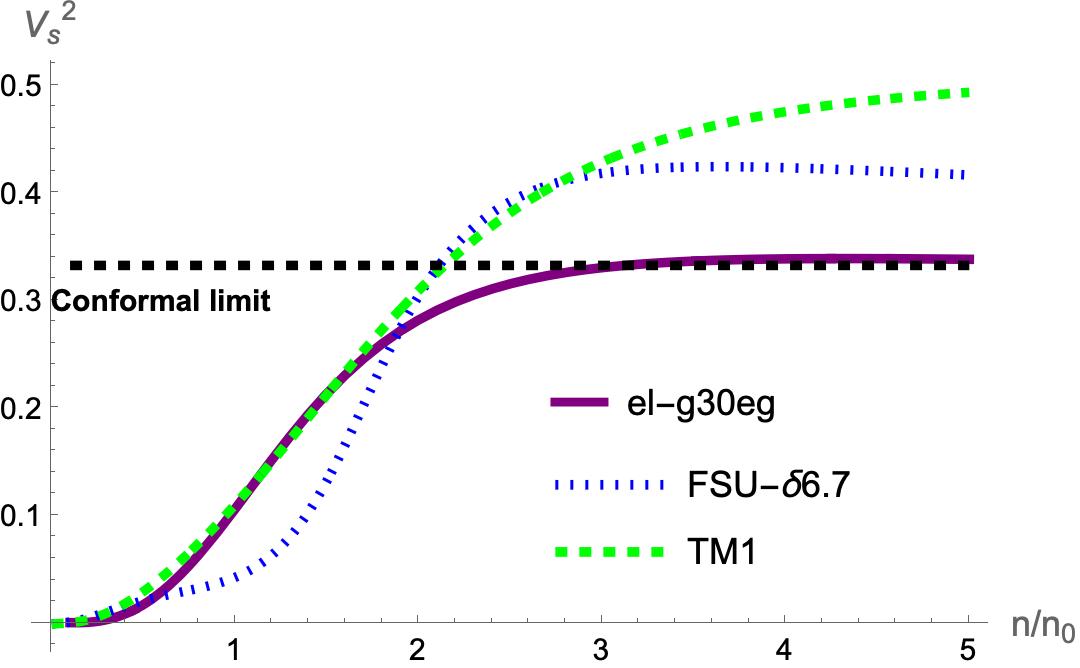}
        \caption{
            Comparison among TM1, FSU-\(\delta 6.7\), and el-g30eg.
            TM1 is from Ref.~\cite{Sugahara:1993wz}.
            FSU-\(\delta 6.7\) is from Ref.~\cite{Li:2022okx}.
        }
        \label{fig:vssub1}
    \end{subfigure}
    \begin{subfigure}[]{0.45\textwidth}
        \includegraphics[width=\textwidth]{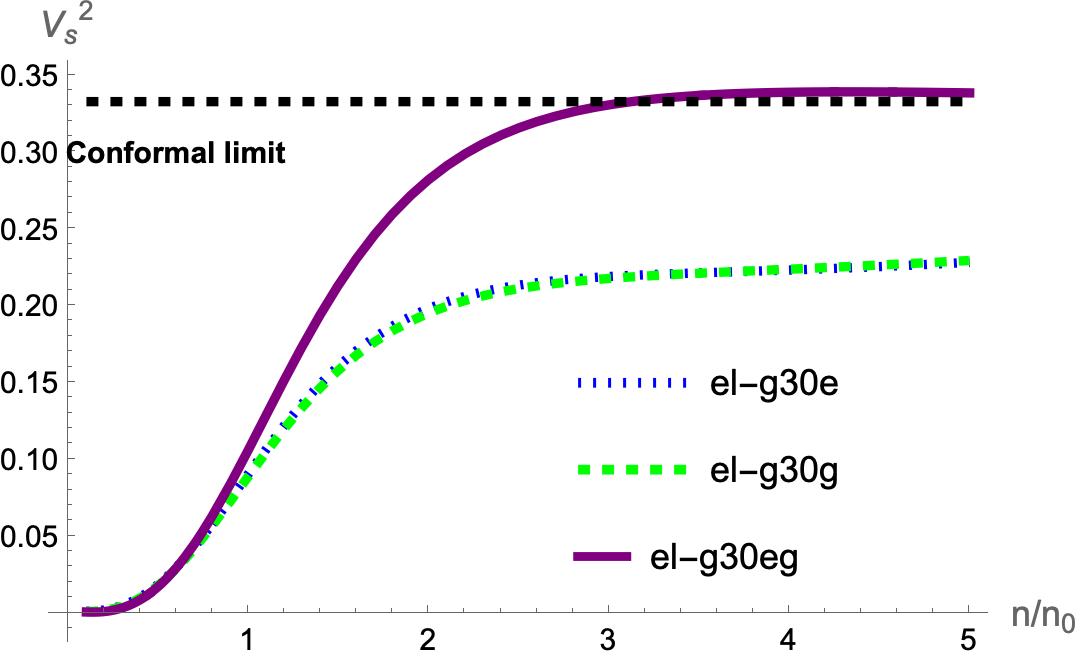}
        \caption{Comparison among cases with different magnitudes of \(e\) and \(g\) flavor structures.}
        \label{fig:vssub2}
    \end{subfigure}
    \begin{subfigure}[]{0.45\textwidth}
        \includegraphics[width=\textwidth]{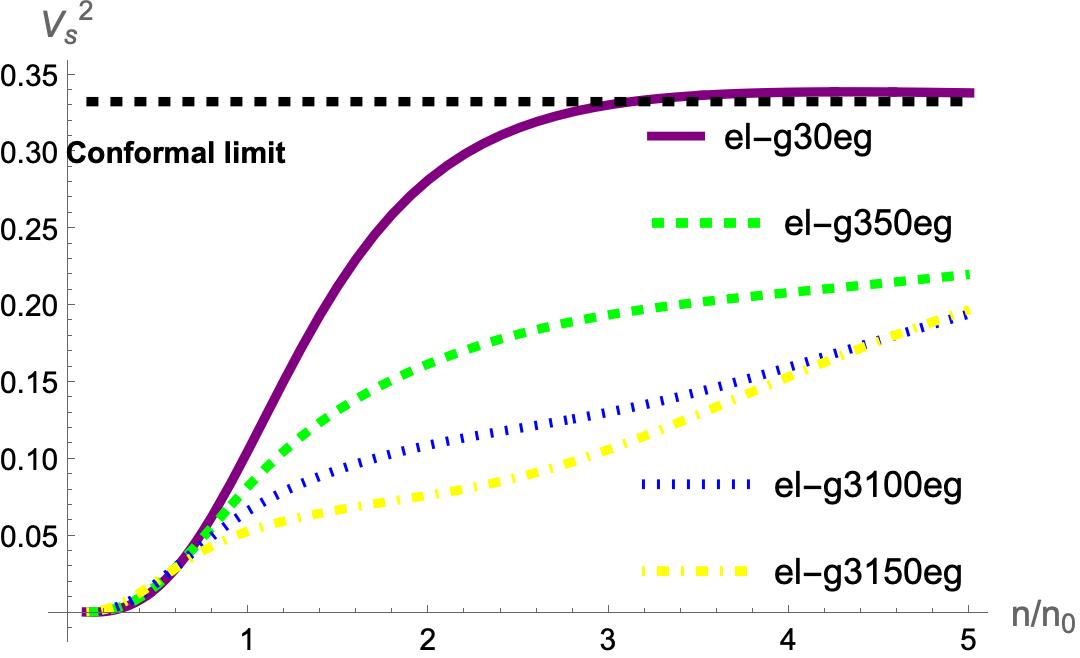}
        \caption{Comparison among cases with different magnitudes of \(\tilde{g}_3\) terms.}
        \label{fig:vssub3}
    \end{subfigure}
    \caption{
        The \(v_s^2\) from different cases.
        The conformal limit is due to the asymptotic freedom of QCD at high energies.
    }
    \label{fig:vs}
\end{figure}

Furthermore, the \(E_{\rm sym}(n)\) across the relevant densities is shown in Fig.~\ref{fig:Esym}.
As shown in Fig.~\ref{fig:Esymsub1}, the \(E_{\rm sym}(n)\) from el-g30eg and FSU-\(\delta\)6.7 is soft at intermediate densities, around \(2n_0\), while stiff at subsaturation densities, around \(2/3n_0\), to meet the constraints of GW170817~\cite{LIGOScientific:2018cki} and the neutron skin thickness of \(^{208}\rm Pb\)~\cite{PREX:2021umo,Reed:2021nqk}.
However, the \(E_{\rm sym}(n)\) from TM1 is almost linearly dependent on the density, which is too stiff at intermediate densities, leading to too-loose NS compared to the observations of GW170817, as shown in Fig.~\ref{fig:MRsub1}.
This highlights the necessity of including the \(a_0\) meson in the NS EOS, resulting in a plateaulike structure of \(E_{\rm sym}(n)\) at intermediate densities. However, a large \(g_{a NN}\) will lead to a decreasing \(E_{\rm sym}(n)\) at high densities, as shown in Fig.~\ref{fig:Esymsub2}, which results in a decrease in the \(M_{\rm max}\) of NSs, as mentioned in Fig.~\ref{fig:MRsub2}.
When comparing the \(E_{\rm sym}(n)\) from different \(\tilde{g}_3\) terms in Fig.~\ref{fig:Esymsub3}, it seems that \(E_{\rm sym}(n)\) is not significantly suppressed by the \(\tilde{g}_3\) term, but the \(M_{\rm max}\) of NSs is even smaller than in the case with a large \(g_{a NN}\).
The reason lies in the SV behavior at high densities, as presented in Fig.~\ref{fig:vs}.
It can be observed that a large \(\tilde{g}_3\) makes \(v_s^2\) softer compared to \(g_{{a_0 NN}}\) suppression, as seen in Figs.~\ref{fig:vssub2} and \ref{fig:vssub3}.
The soft SV behavior, especially when not approaching the conformal limit throughout the densities, predicts a small \(M_{\rm max}\) for NSs, which cannot meet the constraints of MSP J0740+6620.

Now, let us examine the results in Fig.~\ref{fig:MRsub1}:
\begin{itemize}
    \item The SV of TM1 is the stiffest throughout the densities, but the predicted NS is the loosest.
    \item The SV of FSU-\(\delta 6.7\) is the softest at low densities but becomes stiff at intermediate/high densities, exceeding the conformal limit \(v_s^2=1/3\), leading to the most compact NS.
    \item The SV of el-g30eg is similar to TM1 at low densities but approaches the conformal limit at high densities, with NS results consistent with the observations of both GW170817 and MSP J0740+6620.
\end{itemize}
All these indicate that the SV behavior should be stiff enough at high densities to make the \(M_{\rm max}\) of NSs reach at least the observed value of MSP J0740+6620, and the SV behavior should be soft enough at intermediate densities to make the NSs compact enough to meet the constraints of GW170817.
This reflects the subtle relationship between NS structures and SV.

If we only consider the NM properties and the NS structures, both el-g30eg and FSU-\(\delta 6.7\) can meet the phenomenological constraints, with FSU-\(\delta 6.7\) appearing more favored.
However, bELSM reflects the symmetry pattern of QCD at low energies, while the Walecka-type models lack such consideration but introduce possible couplings artificially.
Moreover, considering the SV behavior at high densities, el-g30eg allows the SV to approach the conformal limit. This technically stems from including all possible terms about the related baryons and mesons at the lowest-order, but the deeper reason may be its close relation to QCD.
All the M-R relations in this work are calculated with parameters determined by the NM properties at \(n_0\), but the parameter space and the resulting NS structures can be further improved via statistical analysis, including constraints from NS observations, as discussed in our past work~\cite{Guo:2023mhf}.

\subsection{Constituents of the scalar mesons}

The above discussions suggest possible magnitudes of the two-quark and four-quark constituents of the scalar mesons. We list the percentage of the two-quark component \(p_2=\left|\cos^2\theta\right|\) and four-quark component \(p_4=\left|\sin^2\theta\right|\) defined by Eq.~\eqref{eq:mixing} in Table~\ref{tab:2-4mxing}.
\begin{table}[htb]\small
    \caption{
        The possible percentage of the two-quark and four-quark constituents of the scalar mesons.
    }
    \label{tab:2-4mxing}
    \begin{threeparttable}
        \begin{tabular}{@{}ccccccc}
            \hline
            & \multicolumn{2}{c}{$\sigma$} & & \multicolumn{2}{c}{${a_0}$} \\
            \cline{2-3} \cline{5-6}
            & \(p_4\) & \(p_2\) & & \(p_4\) & \(p_2\) \\
            \hline
            el-g30g & $51.7\%$ & $48.3\%$ & & $77.9\%$ & $22.1\%$ \\
            \hline
            el-g30e & $51.2\%$ & $48.7\%$ & & $77.2\%$ & $22.8\%$ \\
            \hline
            el-g30eg & $54.2\%$ & $45.8\%$ & & $81.5\%$ & $18.5\%$ \\
            \hline
            el-g350eg & $52.4\%$ & $47.6\%$ & & $77.3\%$ & $20.7\%$ \\
            \hline
            el-g3100eg & $49.2\%$ & $50.8\%$ & & $74.4\%$ & $25.6\%$ \\
            \hline
            el-g3150eg & $48.9\%$ & $51.1\%$ & & $74.3\%$ & $25.7\%$ \\
            \hline
        \end{tabular}
	\end{threeparttable}
\end{table}
It is found that the \(\sigma\) meson, a flavor singlet, is composed of two-quark and four-quark constituents in similar proportions, while the \({a_0}\) meson, a flavor octet, is primarily composed of four-quark constituents.
This conclusion is drawn after determining the bare masses of the relevant hadrons in Table~\ref{tab:mass} and the NM properties in Table~\ref{tab:nm}.
This finding is consistent with previous analyses in vacuum~\cite{Fariborz:2007ai}.
When increasing the magnitude of the \(\tilde{g}_3\) term, the four-quark constituent percentages of both the \(\sigma\) and \({a_0}\) mesons decrease to find an optimal parameter set that meets empirical constraints.
Conversely, when the contributions of the \(e\) and \(g\) terms are made similar, the four-quark constituent percentages of both the \(\sigma\) and \({a_0}\) mesons increase.

Regarding the parameter space found in our previous work~\cite{Ma:2023eoz}, the current optimal parameter spaces are more favored, not only for their consistent NM properties at \(n_0\) and reasonable EOS behaviors over a wider density region, as discussed above but also for their prediction of the pion decay constant \(f_{\pi}\).
The pion decay constant can be obtained using the partially conserved axial current (PCAC) relation in the chiral limit (no explicit symmetry breaking due to quark mass)~\cite{Fariborz:2007ai}: \(f_{\pi}=2\sqrt{\alpha^2+\beta^2}\).
The el-g30eg parameter set indicates that \(f_{\pi}\approx134~\rm MeV\), which is consistent with recent lattice QCD results~\cite{FlavourLatticeAveragingGroup:2019iem}, while \(f_{\pi}\sim (60-70)~\rm MeV\) was obtained from the parameter space in our previous work.
This again highlights the importance of the new flavor structures in the phenomenological analysis to achieve a realistic parameter space.

The \(a_0\) meson plays a crucial role in NM studies, particularly for pure neutron matter at high densities~\cite{Kubis:1997ew,Hofmann:2000vz,Liu:2001iz,Chen:2010qx,Zabari:2018tjk,Li:2022okx}. The couplings related to the \(a_0\) meson are traditionally introduced artificially in Walecka-type models and are usually fixed by the properties of NM around \(n_0\).
In this work, to reproduce the rational spectra of mesons and baryons, with respect to the chiral symmetry pattern of QCD at low energies, along with the NM properties at \(n_0\), the couplings related to the \(a_0\) meson are tightly constrained. 

Effectively, we define the coupling through
\be
 \mathcal{L}_{\rm RMF} & = & C_{\sigma{a_0}^2}\sigma a_0^2 + C_{\sigma^2{a_0}^2}\sigma^2 a_0^2.
\ee
Then, according to Table~\ref{tab:para} and Lagrangian~\eqref{eq:Vm}, the corresponding values of the couplings are listed in Table~\ref{tab:delta-sigma} for comparison.
\begin{table}[htb]\small
    \caption[Caption]{
        The values of \(C_{\sigma{a_0}^2}\) and \(C_{\sigma^2{a_0}^2}\) in different models. 
    }
    \label{tab:delta-sigma}
    \begin{threeparttable}
        \begin{tabular}{@{}ccc}
            \hline
            \hline
            & \(C_{\sigma{a_0}^2}\)(MeV) & \(C_{\sigma^2{a_0}^2}\) \\
            \hline
            Zabari-19~\cite{Zabari:2018tjk} & \(\pm1.77\) & \(\pm0.004\) \\
            \hline
            FSU-\(\delta\)6.7~\cite{Li:2022okx} & --- & $2.63$ \\
            \hline
            el-g30g & $-1860$ & $-9.40$ \\
            \hline
            el-g30e & $-1940$ & $-9.71$ \\
            \hline
            el-g30eg & $-1480$ & $-7.70$ \\
            \hline
            el-g350eg & $-1690$ & $-8.75$ \\
            \hline
            el-g3100eg & $-2190$ & $-11.0$ \\
            \hline
            el-g3150eg & $-2160$ & $-11.0$ \\
            \hline
            \hline
        \end{tabular}
    \end{threeparttable}
\end{table}
It can be seen that the magnitude of couplings between \(a_0\) and \(\sigma\) mesons are all large, differing significantly from the results of FSU-\(\delta 6.7\) and Zabari-19, which are used for dense NM studies.

The larger magnitude of the couplings in Table~\ref{tab:delta-sigma} indicates that these multimeson interactions contribute more to the EOS at both saturation density and high densities, and links the behavior of both regions more closely.
This is the reason why we must increase the \(K(n_0)\) to achieve a reasonable NS structures, as shown Fig.~\ref{fig:MRsub3}.
While the parameters from FSU-$\delta$6.7 and Zabari-19 make NM properties at $n_0$ do not change obviously compared to those without couplings between the $a_0$ and $\sigma$ mesons, for the small magnitudes of these couplings, they affect the EOS at high densities much. This leads to the fact that they can handle the NM properties at \(n_0\) and NS structures separately, compared to the bELSM.
However, since the scalar meson couplings in \(\mathcal{L}_{\rm M}\) are determined by \(\alpha\), \(\beta\), \(e_3\), and \(c_4\), which are closely related to the quark constituents of scalar mesons listed in Table~\ref{tab:2-4mxing} and important physical values, such as \(f_{\pi}\).
In contrast, the Walecka-type models, as we are addressing, only introduce possible freedoms and choose operators artificially, which cannot be directly linked to QCD. Therefore, the results in Table~\ref{tab:delta-sigma} highlight the importance of considering the chiral symmetry pattern of QCD at low energies in phenomenological analysis and including operators systematically, especially for phenomenological studies at high densities.

\section{Conclusion}
\label{sec:DO}

In this work, we continued to study NM properties and NS structures by extending the bELSM previousely proposed in Ref.~\cite{Ma:2023eoz} to include a new type of flavor structure which is constructed with an antisymmetric tensor of the corresponding baryon and meson representations.
With this extension, the bELSM at the lowest-order could reproduce the empirical spectra of the scalar mesons and nucleons as well as the NM properties at saturation density. 

Explicitly, we found that without introducing the two-quark terms as in the previous work, the reasonable behavior of nuclear matter at high densities can be well reproduced.
The couplings related to \(\rho\) and \(\omega\) mesons can be separated due to the new flavor structures, allowing the \(E_{\rm sym}\) plateau to appear at the expected density range, around \(2n_0\), instead of below \(n_0\) as in the previous work.
This is consistent with Ref.~\cite{Li:2022okx}, which highlights the importance of such \(E_{\rm sym}\) behavior to meet the constraints of GW170817 and the neutron skin thickness of \(^{208}\rm Pb\) with a unified model.
Besides the phenomena at densities, the pion decay constant is also consistent with recent lattice QCD results~\cite{FlavourLatticeAveragingGroup:2019iem}, with \(f_{\pi}\approx 134~\rm MeV\) in el-g30eg, compared to \(f_{\pi}\sim (60-70)~\rm MeV\) in the previous work.

With this improved lowest-order bELSM and the corresponding parameter space, a further study on the EOS over a wider density region was performed.
Several parameter sets were found to meet the NM properties at \(n_0\).
The differences lie in the magnitude of \(g_{{a_0 NN}}\) and the four-vector meson coupling, \(\tilde{g}_3\).
It was found that small \(g_{{a_0 NN}}\) and \(\tilde{g}_3\) lead to more compact NSs, consistent with the constraints of MSP J0740+6620 and GW170817.
The influences of the EOS behaviors on NS M-R relations were also discussed: a soft \(E_{\rm sym}\) at intermediate densities, with a stiff behavior at subsaturation densities, is necessary for the compactness of NSs to meet the GW170817 constraints, and a stiff SV behavior at high densities is necessary for the \(M_{\rm max}\) of NSs to meet the MSP J0740+6620 constraints.
Besides the well-produced NS structures from el-g30eg and FSU-\(\delta 6.7\), the SV behavior of el-g30eg is more favored, as its SV behavior can approach the conformal limit at high densities, expected for the fact that QCD is asymptotically free at high energies.

The percentages of two-quark and four-quark constituents making up the scalar mesons were also calculated.
The \(\sigma\) meson was found to be composed of two-quark and four-quark constituents in similar proportions, while the \(a_0\) meson is mainly composed of four-quark constituents in the current optimal parameter spaces.
This is consistent with previous analyses in vacuum~\cite{Fariborz:2007ai}.
The variation of the constituents due to different magnitudes of \(g_{{a_0 NN}}\) and four-vector meson couplings were also provided for future studies on scalar mesons and the macroscopic properties of compact objects in the universe.
However, the quark constituents make the couplings related to \(a_0\), both types: \(g_{{a_0 NN}}\) and \(C_{\sigma^i{a_0}^2}\), {with \(i=1,\ 2\)}, different from the results of FSU-\(\delta 6.7\) and Zabari-19, which are used for the symmetry energy at high densities.
This again signifies the importance of considering the chiral symmetry pattern of QCD at low energies in phenomenological analyses beyond the saturation density.

In summary, the lowest-lying isoscalar and isovector scalar and vector mesons are all included within a unified model, bELSM, which also respects the chiral symmetry pattern of QCD at low energies.
The NM properties and NS structures are systematically studied. Our recent works~\cite{Guo:2023mhf,Ma:2023eoz,Guo:2024nzi,zhang2024peaksoundvelocityscale,Zhang:2024iye} focus on the relationships between the QCD symmetry pattern and compact objects in the universe.
In the near future, improved AI analysis on a general Walecka-type model with the inclusion of all types of lowest-lying mesons and baryons will be available, which will help further investigate bELSM at {finite} densities due to its numerical complexity.
Additionally, studies based on bELSM with the inclusion of strange freedoms are underway to address the properties of compact objects with hyperons.
Meanwhile, the Hartree-Fock approach with chiral effective field theories will also be initially completed to study the EOS of NM, including quantum effects, which are ignored in the current RMF approximation.

\acknowledgments

The work of Y.~L. M. is supported in part by the National Science Foundation of China (NSFC) under Grant No. 12347103, the National Key R\&D Program of China under Grant No. 2021YFC2202900 and Gusu Talent Innovation Program under Grant No. ZXL2024363.

\appendix

\section{The lowest-order bELSM}
\label{app:LbELSM}

In this Appendix, we provide details of the bELSM at the lowest-order before the RMF approximation. In the model, the scalar mesons $S$ and pseudoscalar mesons $P$ are combined into a compact form \(\Phi=S+iP\):
\begin{widetext}
\begin{equation}
    \Phi=S+i P=\left(\begin{array}{ccc}
        \frac{\left(\sigma_N+a_0^0\right)+i\left(\eta_N+\pi^0\right)}{\sqrt{2}} & a_0^{+}+i \pi^{+} & K_S^{+}+i K^{+} \\
        a_0^{-}+i \pi^{-} & \frac{\left(\sigma_N-a_0^0\right)+i\left(\eta_N-\pi^0\right)}{\sqrt{2}} & K_S^0+i K^0 \\
        K_S^{-}+i K^{-} & \bar{K}_S^0+i \bar{K}^0 & \sigma_S+i \eta_S
    \end{array}\right)\ .
\end{equation}
The representations of the two-quark and four-quark (pseudo-)scalar configurations are similar, but with "prime" and "hat-prime" indices, as defined in the main text.
For the vector and axial-vector mesons, they are expressed as,
\begin{equation}
    (R, L)^\mu=V^\mu \pm A^\mu=\frac{1}{\sqrt{2}}\left(\begin{array}{ccc}
        \frac{\omega_N^\mu+\rho^{\mu 0}}{\sqrt{2}} \pm \frac{f_{1 N}^\mu+a_1^{\mu 0}}{\sqrt{2}} & \rho^{\mu+} \pm a_1^{\mu+} & K^{* \mu+} \pm K_1^{\mu+} \\
        \rho^{\mu-} \pm a_1^{\mu-} & \frac{\omega_N^\mu-\rho^{\mu 0}}{\sqrt{2}} \pm \frac{f_{1 N}^\mu-a_1^{\mu 0}}{} & K^{* \mu 0} \pm K_1^{\mu 0} \\
        K^{* \mu-} \pm K_1^{\mu-} & \bar{K}^{* \mu 0} \pm \bar{K}_1^{\mu 0} & \omega_S^\mu \pm f_{1 S}^\mu
    \end{array}\right)\ ,
\end{equation}
where \(L(R)^{\mu}\) stands for the left-(right)-handed vector mesons, and \(V(A)^{\mu}\) denotes the vector(axial-vector) mesons.
\end{widetext}
The baryon octet has expression:
\begin{equation}
    B =\left(\begin{array}{ccc}
        \frac{\Lambda}{\sqrt{6}}+\frac{\Sigma^0}{\sqrt{2}} & \Sigma^{+} & p \\
        \Sigma^{-} & \frac{\Lambda}{\sqrt{6}}-\frac{\Sigma^0}{\sqrt{2}} & n \\
        \Xi^{-} & \Xi^0 & -\frac{2 \Lambda}{\sqrt{6}}
    \end{array}\right)\ .
\end{equation}

The chiral symmetry transformations for (pseudo)scalar configurations are given by~\cite{Fariborz:2005gm,Fariborz:2007ai,Fariborz:2009cq}:
\begin{equation}
    \Phi' \rightarrow g_{\mathrm{L}} \Phi' g_{\mathrm{R}}^{\dagger}, \quad \hat{\Phi}' \rightarrow g_{\mathrm{L}} \hat{\Phi}' g_{\mathrm{R}}^{\dagger}\ ,
\end{equation}
where \(g_{\mathrm{L}, \mathrm{R}} \in \mathrm{SU}(3)_{\mathrm{L}, \mathrm{R}}\).
They remain unchanged for \(\rm U(1)_V\) transformation. But for \(\rm U(1)_A\) transformation, they transform as,
\begin{equation}
    \Phi' \rightarrow e^{2 i \nu} \Phi', \quad \hat{\Phi}' \rightarrow e^{-4 i \nu} \hat{\Phi}'\ ,
\end{equation}
with \(\nu\) being the chiral angle, and \(q \bar{q}\) and \((q q)(\bar{q} \bar{q})\) configurations are assumed for two-quark and four-quark components, respectively.
For the vector and axial-vector mesons, their chiral transformations are~\cite{Parganlija:2012fy},
\begin{equation}
    L_\mu \rightarrow g_{\mathrm{L}} L_\mu g_{\mathrm{L}}^{\dagger}, \quad R_\mu \rightarrow g_{\mathrm{R}} R_\mu g_{\mathrm{R}}^{\dagger}\ ,
\end{equation}
with \(\bar{q}_{\mathrm{L(R)}} \gamma_\mu q_{\mathrm{L(R)}}\) configuration assumed. They are both invariant for both \(\rm U(1)_V\) and \(\rm U(1)_A\) transformations.

In this work, the diquark approximation for baryons is adopted~\cite{thesis2015Olbrich,Olbrich:2015gln}.
The baryons are denoted as \(N_{\mathrm{R}}^{(\mathrm{RR})}\), \(N_{\mathrm{L}}^{(\mathrm{LL})}\), \(N_{\mathrm{L}}^{(\mathrm{RR})}\), and \(N_{\mathrm{R}}^{(\mathrm{LL})}\), where the superscripts represent the quark configuration and the subscripts represent the diquark configuration.
The chiral transformations for baryons are
\be
N_{\mathrm{R}}^{(\mathrm{RR})} & \rightarrow & g_{\mathrm{R}} N_{\mathrm{R}}^{(\mathrm{RR})} g_{\mathrm{R}}^{\dagger}\ ,\quad N_{\mathrm{L}}^{(\mathrm{RR})} \rightarrow g_{\mathrm{L}} N_{\mathrm{L}}^{(\mathrm{RR})} g_{\mathrm{R}}^{\dagger}\ , \nm\\
N_{\mathrm{R}}^{(\mathrm{LL})} & \rightarrow & g_{\mathrm{R}} N_{\mathrm{R}}^{(\mathrm{LL})} g_{\mathrm{L}}^{\dagger}\ ,\quad N_{\mathrm{L}}^{(\mathrm{LL})} \rightarrow g_{\mathrm{L}} N_{\mathrm{L}}^{(\mathrm{LL})} g_{\mathrm{L}}^{\dagger}\ .
\ee
And the \(\rm U(1)_{A}\) transformations are
\be
N_{\mathrm{R}}^{(\mathrm{RR})} & \rightarrow & e^{-3 i v} N_{\mathrm{R}}^{(\mathrm{RR})}\ , \quad N_{\mathrm{L}}^{(\mathrm{RR})} \rightarrow e^{-i v} N_{\mathrm{L}}^{(\mathrm{RR})}\ , \nm\\
N_{\mathrm{R}}^{(\mathrm{LL})} & \rightarrow & e^{i v} N_{\mathrm{R}}^{(\mathrm{LL})}\ , \quad N_{\mathrm{L}}^{(\mathrm{LL})} \rightarrow e^{3 i v} N_{\mathrm{L}}^{(\mathrm{LL})}\ ,
\ee
with invariant under \(\rm U(1)_{V}\) transformation.
The physical baryons are defined as:
\begin{equation}
    \label{eq:physbaryon}
    N_{\mathrm{R}, \mathrm{~L}}^{(\mathrm{RR})}=\frac{1}{\sqrt{2}} \frac{1 \pm \gamma_5}{2} B\ ,\quad N_{\mathrm{R}, \mathrm{~L}}^{(\mathrm{LL})}=-\frac{1}{\sqrt{2}} \frac{1 \pm \gamma_5}{2} B\ ,
\end{equation}
with excited state \(N^*\) being neglected since it is irrelevant to the present work.

Before constructing the full Lagrangian, a series of power counting rules is needed to organize the effective operators:
\begin{itemize}
    \item The operators are limited to dimension-4, as higher-dimensional operators are suppressed by the cutoff scale.
    \item The quark number of an operator is limited to 8~\cite{Fariborz:2005gm,Fariborz:2007ai,Fariborz:2009cq}, as operators with more quarks are suppressed in the effective description of quark lines.
    \item The number of flavor space traces is limited to 1, since more trace terms are suppression by \(N_c\)~\cite{Parganlija:2010fz}.
    \item The explicit symmetry breaking caused by quark mass is treated as a perturbation and is ignored in the current work.
\end{itemize}

Consequently, the full Lagrangian at the lowest-order is 
\be
\mathcal{L}_{\rm LO}=\mathcal{L}_{\rm M}+\mathcal{L}_{\rm V}+\mathcal{L}_{\rm B}.
\ee
The (pseudo)scalar meson part is~\cite{Fariborz:2007ai}
\begin{widetext}
\be
\mathcal{L}_{\rm M} & = & \frac{1}{2} \operatorname{Tr}\left(\partial_\mu \Phi' \partial^\mu \Phi'^{\dagger}\right)+\frac{1}{2} \operatorname{Tr}\left(\partial_\mu \hat{\Phi}' \partial^\mu\hat{\Phi}'^{\dagger} \right)+c_2 \operatorname{Tr}\left(\Phi'\Phi'^{\dagger}\right)-c_4 \operatorname{Tr}\left(\Phi' \Phi'^{\dagger} \Phi' \Phi'^{\dagger}\right)- d_2 \operatorname{Tr}\left(\hat{\Phi}' \hat{\Phi}'^{\dagger}\right)\nonumber\\
& &{} - e_3\left(\epsilon_{a b c} \epsilon^{d e f} \Phi_d^{\prime a}\Phi_e^{\prime b} \hat{\Phi}_f^{\prime c}+{\rm h.c.}\right)- c_3\left[\gamma_1 \ln \left(\frac{\operatorname{det} \Phi'}{\operatorname{det} {\Phi}'^{\dagger}}\right)+\left(1-\gamma_1\right) \ln \left(\frac{\operatorname{Tr}\left(\Phi' \hat{\Phi}'^{\dagger}\right)}{\operatorname{Tr}\left(\hat{\Phi}' \Phi'^{\dagger}\right)}\right)\right]^2\ ,
\ee
where the \(c_3\) term accounts for \(\rm U(1)_A\) anomaly at low energies.
The (axial-)vector meson part is written as:
\be
\mathcal{L}_{\rm V} & = &{} -\frac{1}{8}{\rm Tr}\left({L}_{\mu \nu} {L}^{\mu \nu}+{R}_{\mu \nu} {R}^{\mu \nu}\right) \nm\\
& &{} + g_1\left[{\rm Tr}\left(\partial_{\nu}{R}_{\mu}{R}^{\mu}{R}^{\nu}\right)+{\rm Tr}\left(\partial_{\nu}{L}_{\mu}{L}^{\mu}{L}^{\nu}\right)\right] + g_2\left[{\rm Tr}\left(\partial_{\nu}{R}_{\mu}{R}^{\nu}{R}^{\mu}\right)+{\rm Tr}\left(\partial_{\nu}{L}_{\mu}{L}^{\nu}{L}^{\mu}\right)\right] \nm\\
& &{} + g_3\left[{\rm Tr}\left({R}_{\mu}{R}^{\mu}{R}_{\nu}{R}^{\nu}\right)+{\rm Tr}\left({L}_{\mu}{L}^{\mu}{L}_{\nu}{L}^{\nu}\right)\right] + g_4\left[{\rm Tr}\left({R}_{\mu}{R}^{\nu}{R}_{\mu}{R}^{\nu}\right)+{\rm Tr}\left({L}_{\mu}{L}^{\nu}{L}_{\mu}{L}^{\nu}\right)\right]\nonumber\\
& &{} + h_1\left[{\rm Tr}\left(\Phi^{\prime\dagger}\Phi^{\prime}R_{\mu}R^{\mu}\right)+{\rm Tr}\left(\Phi^{\prime}\Phi^{\prime\dagger}L_{\mu}L^{\mu}\right)\right] + h_2\left[{\rm Tr}\left(L^{\mu}\Phi'R_{\mu}\Phi^{\prime\dagger}\right)\right]+h_3\left[{\rm Tr}\left(\Phi^{\prime\dagger}\partial_{\mu}\Phi'R^{\mu}\right)+{\rm Tr}\left(\Phi^{\prime}\partial_{\mu}\Phi^{\prime\dagger}L^{\mu}\right)\right] \nonumber\\
& &{} + \frac{a_1}{2}\epsilon_{abc}\epsilon^{def}\left[\left(R_{\mu}\right)_{ad}\left(R_{\nu}\right)_{be}\left(\Phi^{\prime\dagger}\Phi\right)_{cf}+\left(L_{\mu}\right)_{ad}\left(L_{\nu}\right)_{be}\left(\Phi\Phi^{\prime\dagger}\right)_{cf}\right] \nonumber\\
& &{} + \frac{a_2}{2}\epsilon_{abc}\epsilon^{def}\left[\left(R_{\mu}\right)_{ad}\left(R_{\nu}\right)_{be}\left(R^{\mu}R^{\nu}\right)_{cf}+\left(L_{\mu}\right)_{ad}\left(L_{\nu}\right)_{be}\left(L^{\mu}L^{\nu}\right)_{cf}\right] \nonumber\\
& & {} + \frac{a_3}{2}\epsilon_{abc}\epsilon^{def}\left[\left(R_{\mu}\right)_{ad}\left(R^{\mu}\right)_{be}\left(R^{\nu}R_{\nu}\right)_{cf}+\left(L_{\mu}\right)_{ad}\left(L^{\mu}\right)_{be}\left(L^{\nu}L_{\nu}\right)_{cf}\right] \nonumber\\
& &{} + \frac{a_4}{2}\epsilon_{abc}\epsilon^{def}\left[\left(R_{\mu}\right)_{ad}\left(R^{\nu}\right)_{be}\left(\partial^{\mu}R_{\nu}\right)_{cf}+\left(L_{\mu}\right)_{ad}\left(L^{\nu}\right)_{be}\left(\partial^{\mu}L_{\nu}\right)_{cf}\right]\ .
\ee
The baryon part has expression
\be
\mathcal{L}_{\rm B} & = & {\rm Tr}\left(\bar{N}_{\rm R}^{\left(\rm RR\right)}i\slashed{\partial}N_{\rm R}^{\left(\rm RR\right)}+\bar{N}_{\rm L}^{\left(\rm RR\right)}i\slashed{\partial}N_{\rm L}^{\left(\rm RR\right)}+\bar{N}_{\rm R}^{\left(\rm LL\right)}i\slashed{\partial}N_{\rm R}^{\left(\rm LL\right)}+\bar{N}_{\rm L}^{\left(\rm LL\right)}i\slashed{\partial}N_{\rm L}^{\left(\rm LL\right)}\right) \nonumber\\
& &{} + n_1{\rm Tr}\left(\bar{N}_{\rm R}^{\left(\rm RR\right)}\slashed{R}N_{\rm R}^{\left(\rm RR\right)}+\bar{N}_{\rm L}^{\left(\rm LL\right)}\slashed{L}N_{\rm L}^{\left(\rm LL\right)}\right)+n_2{\rm Tr}\left(\bar{N}_{\rm R}^{\left(\rm LL\right)}\slashed{R}N_{\rm R}^{\left(\rm LL\right)}+\bar{N}_{\rm L}^{\left(\rm RR\right)}\slashed{L}N_{\rm L}^{\left(\rm RR\right)}\right) \nonumber\\
& &{} +n_3{\rm Tr}\left(\bar{N}_{\rm R}^{\left(\rm RR\right)}\gamma_{\mu}N_{\rm R}^{\left(\rm RR\right)}R^{\mu}+\bar{N}_{\rm L}^{\left(\rm LL\right)}\gamma_{\mu}N_{\rm L}^{\left(\rm RR\right)}L^{\mu}\right)+n_4\left(\bar{N}_{\rm R}^{\left(\rm LL\right)}\gamma_{\mu}N_{\rm R}^{\left(\rm LL\right)}L^{\mu}+\bar{N}_{\rm L}^{\left(\rm RR\right)}\gamma_{\mu}N_{\rm L}^{\left(\rm RR\right)}R^{\mu}\right) \nonumber\\
& &{} + n_{a}\epsilon_{abc}\epsilon^{def}\left[\left(\bar{N}_{\rm R}^{\left(\rm RR\right)}\right)_{ad}\gamma_{\mu}\left(N_{\rm R}^{\left(\rm RR\right)}\right)_{be}\left(R^{\mu}\right)_{cf}+\left(\bar{N}_{\rm L}^{\left(\rm LL\right)}\right)_{ad}\gamma_{\mu}\left(N_{\rm L}^{\left(\rm LL\right)}\right)_{be}\left(L^{\mu}\right)_{cf}\right] \nonumber\\
& &{} + n_g{\rm Tr}\left[\bar{N}_{\rm R}^{\left(\rm RR\right)}\Phi^{\prime\dagger}N_{\rm L}^{\left(\rm RR\right)}+\bar{N}_{\rm L}^{\left(\rm LL\right)}\Phi^{\prime}N_{\rm R}^{\left(\rm LL\right)}+{\rm h.c.}\right] \nonumber\\
& &{} + n_e\epsilon_{abc}\epsilon^{def}\left[\left(\bar{N}_{\rm R}^{\left(\rm RR\right)}\right)_{ad}\left(\Phi^{\prime\dagger}\right)_{be}\left(N_{\rm L}^{\left(\rm RR\right)}\right)_{cf}+\left(\bar{N}_{\rm L}^{\left(\rm LL\right)}\right)_{ad}\left(\Phi^{\prime}\right)_{be}\left(N_{\rm R}^{\left(\rm LL\right)}\right)_{cf}+{\rm h.c.}\right]\ .
\ee
And after substituting Eq.~\eqref{eq:physbaryon} into the Lagrangian, the baryon part can be rewritten as
\be
\mathcal{L}_{\rm B} & = & {\rm Tr}\left(\bar{B}i\slashed{\partial}B\right)+c{\rm Tr}\left(\bar{B}\gamma_{\mu}V^{\mu}B+\bar{B}\gamma_{\mu}\gamma_5A^{\mu}B\right)+{c'{\rm Tr}\left(\bar{B}\gamma_{\mu}B V^{\mu}\right)+c'_A\left(\bar{B}\gamma_{\mu}\gamma_5 B A^{\mu}\right)} \nonumber\\
& &{} + h\epsilon_{abc}\epsilon^{def}\left[\left(\bar{B}\right)_{ad}\gamma_{\mu}\left(B\right)_{be}\left(V^{\mu}\right)_{cf}+\left(\bar{B}\right)_{ad}\gamma_{\mu}\gamma_5\left(B\right)_{be}\left(A^{\mu}\right)_{cf}\right] \nonumber\\
& &{} - \frac{g}{2}{\rm Tr}\left[\bar{B}\left(\Phi^{\prime}+\Phi^{\prime\dagger}\right)B+\bar{B}\gamma_5\left(\Phi^{\prime}-\Phi^{\prime\dagger}\right)B\right] \nonumber\\
& &{} - \frac{e}{2}\epsilon_{abc}\epsilon^{def}\left[\left(\bar{B}\right)_{ad}\left(\Phi^{\prime}+\Phi^{\prime\dagger}\right)_{be}\left(B\right)_{cf}+\left(\bar{B}\right)_{ad}\gamma_5\left(\Phi^{\prime}-\Phi^{\prime\dagger}\right)_{be}\left(B\right)_{cf}\right]\ ,
\ee
where \(c=(n_1+n_2)/2\), {\(c'=(n_3+n_4)/2\), \(c'_A=(n_3-n_4)/2\)}, \(h=n_{a}/2\), \(g=-n_{g}\) and \(e=-n_{e}\).
\end{widetext}
	
\section{Approximations in the current work}
\label{app:app}

In the present study of NM properties, we take RMF, which is a classical approximation, where meson fields are treated as potentials induced by matter fields, such as nucleons in the current work.
Additionally, the system composed of these baryons is considered to be infinite, homogeneous, and static, so that any quantum numbers and currents are averaged to zero.
This results in the approximations
\be
\langle\omega^{\mu}\rangle & = & \langle\omega^0\rangle\delta^{\mu,0}=\omega\delta^{\mu,0}\ ,\nonumber\\
\langle\rho_i^{\mu}\rangle & = & \langle\rho_0^{0}\rangle\delta_{i,0}\delta^{\mu,0}=\rho\delta_{i,0}\delta^{\mu,0}\ ,\nonumber\\
\langle\partial_{\mu}\sigma\rangle & = & \partial_{\mu}\langle\sigma\rangle=0\ ,\nonumber\\
\langle\pi^{i}\rangle & = & 0\ ,
\ee
and other related meson fields are also treated similarly.
For instance, \(K\) mesons are excluded due to their nonzero strange quantum number.

Mesons heavier than \(1~\rm GeV\), such as \(a'_0\) and \(\omega_S\), are ignored since their contributions are suppressed at the current working scale.
This leads to the following parameter choices in the current work: \(\tilde{g}_3=2(g_3+g_4)\), \(\tilde{h}_2=2h_1+h_2\), while the other parameters remain the same as the original ones after the approximations.
Details on how to solve the RMF equations of motion to obtain the equation of state after these approximations are provided in our other work~\cite{Guo:2024nzi}.

\bibliography{draft}

\end{document}